\documentclass[twocolumn,preprintnumbers,amsmath,amssymb,superscriptaddress, bibnotes]{revtex4}

\usepackage{graphicx}
\usepackage{dcolumn}
\usepackage{bm}
\usepackage{xspace}
\usepackage{braket}

\newcommand{\EF}{$E_\mathrm{F}$\xspace}
\newcommand{\Uf}{$\mathrm{U}~5f$\xspace}
\newcommand{\orb}[2]{$\mathrm{ #1 } ~ #2 $\xspace}
\newcommand{\hn}[1]{$h\nu #1~\mathrm{eV}$\xspace}
\newcommand{\EB}[1]{$E_{\mathrm{B}} #1~\mathrm{eV}$\xspace}

\newcommand{\etal}{\textit{et al.}\xspace}

\newcommand{\UAl}{$\mathrm{UAl_3}$\xspace}

\newcommand{\UPdAl}{$\mathrm{UPd_{2}Al_{3}}$\xspace}

\newcommand{\UGa}{$\mathrm{UGa_2}$\xspace}

\newcommand{\Udf}{$\mathrm{U}~4d-5f$\xspace}

\begin{document}
\draft
\preprint{HEP/123-qed}

\title{Manifestation of electron correlation effect in $\mathrm{U}~5f$ states of uranium compounds revealed by $\mathrm{U}~4d-5f$ resonant photoemission spectroscopy} 

\author{Shin-ichi~Fujimori}
\affiliation{Materials Sciences Research Center, Japan Atomic Energy Agency, Sayo, Hyogo 679-5148, Japan}

\author{Masaharu~Kobata}
\affiliation{Materials Sciences Research Center, Japan Atomic Energy Agency, Sayo, Hyogo 679-5148, Japan}

\author{Yukiharu~Takeda}
\affiliation{Materials Sciences Research Center, Japan Atomic Energy Agency, Sayo, Hyogo 679-5148, Japan}

\author{Tetsuo~Okane}
\affiliation{Materials Sciences Research Center, Japan Atomic Energy Agency, Sayo, Hyogo 679-5148, Japan}

\author{Yuji~Saitoh}
\affiliation{Materials Sciences Research Center, Japan Atomic Energy Agency, Sayo, Hyogo 679-5148, Japan}

\author{Atsushi~Fujimori}
\affiliation{Materials Sciences Research Center, Japan Atomic Energy Agency, Sayo, Hyogo 679-5148, Japan}
\affiliation{Department of Physics, University of Tokyo, Hongo, Tokyo 113-0033, Japan}

\author{Hiroshi~Yamagami}
\affiliation{Materials Sciences Research Center, Japan Atomic Energy Agency, Sayo, Hyogo 679-5148, Japan}
\affiliation{Department of Physics, Faculty of Science, Kyoto Sangyo University, Kyoto 603-8555, Japan}

\author{Yoshinori~Haga}
\affiliation{Advanced Science Research Center, Japan Atomic Energy Agency, Tokai, Ibaraki 319-1195, Japan}

\author{Etsuji~Yamamoto}
\affiliation{Advanced Science Research Center, Japan Atomic Energy Agency, Tokai, Ibaraki 319-1195, Japan}

\author{Yoshichika~\=Onuki}
\affiliation{Faculty of Science, University of the Ryukyus, Nishihara, Okinawa 903-0213, Japan}

\date{\today}

\begin{abstract}
We have elucidated the nature of the electron correlation effect in uranium compounds by imaging the partial $\mathrm{U}~5f$ density of states (pDOS) of typical itinerant, localized, and heavy fermion uranium compounds by using the $\mathrm{U}~4d-5f$ resonant photoemission spectroscopy.
Obtained $\mathrm{U}~5f$ pDOS exhibit a systematic trend depending on the physical properties of compounds.
The coherent peak at the Fermi level can be described by the band-structure calculation, but an incoherent peak emerges on the higher binding energy side ($\lesssim 1~\mathrm{eV}$) in the \Uf pDOS of localized and heavy fermion compounds.
As the $\mathrm{U}~5f$ state is more localized, the intensity of the incoherent peak is enhanced and its energy position is shifted to higher binding energy.
These behaviors are consistent with the prediction of the Mott metal-insulator transition, suggesting that the Hubbard-$U$ type mechanism takes an essential role in the $5f$ electronic structure of actinide materials.
\end{abstract}

\maketitle
\narrowtext
\section{INTRODUCTION}
The electron correlation effect gives rise to an abundant variety of physical properties particularly in the $d$- and $f$-based materials.
Among this class of materials, the uranium-based compounds share a unique position due to the interplay between their magnetic and superconducting properties.
Particularly, the coexistence of a large magnetic moment and unconventional superconductivity is the most significant aspect of uranium-based compounds \cite{4f5f}.
These distinctive behaviors are due to the strongly correlated \Uf states which are located at the boundary between simple localized and itinerant pictures of electrons.
To understand the origin of these remarkable physical properties, it is essential to unveil the \Uf electronic structures.

Resonant photoemission (RPES) is a powerful experimental tool which is capable of identifying the contribution from a specific atomic orbital in the valence band spectra \cite{Allen_review}.
It has been applied to strongly correlated $d$- and $f$-electron materials, and their $d$ and $f$ partial density of states (pDOS) have been obtained experimentally \cite{Allen_review,Laubschat_review1, Laubschat_review2, SF_review_JPCM}.
For Ce-based compounds, the \orb{Ce}{4d} (\hn{\sim 122}) and the \orb{Ce}{3d} (\hn{\sim 881}) absorption edges have been frequently utilized to image their \orb{Ce}{4d} pDOS.
In the case of uranium compounds, the \orb{U}{5d} absorption edge (\hn{\sim 108}) has been utilized to obtain the \Uf pDOS
\cite{U5d5fRPES}.
However, the mean free path of photoexcited valence electrons at these photon energies has the shortest value ($\lambda \lesssim 5~\mathrm{\AA}$), and the spectra are dominated by the information of less than one unit cell from the surface where the \Uf electrons are much localized than those in the bulk \cite{UPd2Al3_5d5fRPES}.
Although an enhanced bulk sensitivity of more than 15~\AA \xspace is expected at the $\mathrm{U}~4d$ absorption edge (\hn{=736}), the absence of the resonance enhancement of \Uf signals at this absorption edge was reported by Allen \etal \cite{RPES_Allen}.
On the other hand, Tobin \etal recently reported a finite \Udf resonance enhancement in the resonant inverse photoemission (RIPES) spectra of $\mathrm{UO}_2$ \cite{UO2_RIPES}.
The enhancement factor is about 2 which is more than one order smaller than the values of \orb{U}{5d-5f} or \orb{Ce}{3d-4f} RPESs.
Nevertheless, this finite enhancement is enough for the identification of the contributions from the unoccupied \Uf states among other orbitals.

In the present study, we have measured the photon energy dependence of the photoemission spectra of uranium compounds at the \orb{U}{4d} absorption energy very precisely, and discovered the finite enhancement of \Uf signals of about 15--20 \% at the edge.
We have utilized this enhancement to image the bulk \Uf pDOS of some uranium compounds, and unveiled the nature of the electron correlation effect in these compounds.

We selected \UAl, \UGa, and \UPdAl as typical itinerant, localized, and heavy fermion compounds, respectively.
\UAl is a spin-fluctuation system with itinerant \Uf state, and its band structure and Fermi surface are essentially described by the band-structure calculation \cite{UAl3_dHvA}.
On the other hand, \UGa is a prototypical \Uf localized compound \cite{UGa2_CEF,UGa2_calc} which undergoes a ferromagnetic phase below a Curie temperature of $T_\mathrm{C} = 125~\mathrm{K}$.
The heavy fermion superconductor \UPdAl is characterized by the large specific heat coefficient of $\gamma = 210~\mathrm{mJ}/\mathrm{molK^2}$ \cite{UPd2Al3_Geibel}.
It undergoes an antiferromagnetic phase below the Neel temperature of $T_\mathrm{N}=14~\mathrm{K}$ and superconducting phase below $T_\mathrm{SC}= 2~\mathrm{K}$.
Although its overall band structure can be described by the band-structure calculation, the electronic structure in the vicinity of \EF is modified due to the electron correlation effect \cite{UPd2Al3_ARPES1, UPd2Al3_ARPES2, UM2Al3_ARPES, SF_review_JPSJ}.

\begin{figure}[t]
	\includegraphics[scale=0.5]{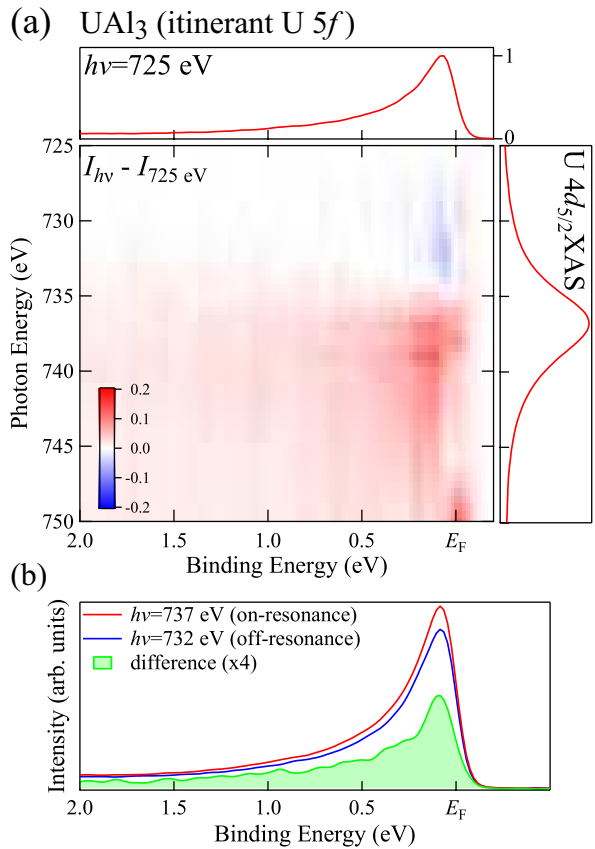}
	\caption{(Online color)
		RPES spectra of \UAl.
		(a) Density plot of RPES spectra together with the \orb{U}{4d_{5/2}} XAS spectrum.
		(b) On- and off-resonance spectra measured at $h \nu = 737$ and $732~\mathrm{eV}$, respectively, and the corresponding difference spectrum.
	}
\label{UAl3_RPES}
\end{figure}
\section{EXPERIMENTAL PROCEDURES}
Photoemission experiments were performed at the soft X-ray beamline BL23SU of SPring-8 \cite{BL23SU2}.
The overall energy resolution at \hn{=720-780} was about $100-130~\mathrm{meV}$.
The on- and off-resonance photon energies were chosen for each compound to minimize the influences of the contributions from ligand states.
Clean sample surfaces were obtained by cleaving high-quality single crystals {\it in situ} under ultra-high vacuum condition.
The sample temperature was kept at $20~\mathrm{K}$ during the measurements for all compounds, and \UAl and \UPdAl were in the paramagnetic phase whereas \UGa was in the ferromagnetic phase.
It should be noted that no recognizable changes were observed between the spectra of \UGa measured above and below $T_\mathrm{C}$.
To evaluate the photon flux on the sample surface, we monitored the photoemission intensities of shallow core-level spectra of ligand atoms, and the photon energy dependencies of their photoionization cross sections were also taken into account by referring the values from the atomic calculation \cite{Atomic}.

\section{RESULTS and DISCUSSION}
\subsection{Itinerant compound \UAl}
Figure~\ref{UAl3_RPES} shows the \Udf RPES spectra and the \orb{U}{4d_{5/2}} X-ray absorption spectroscopy (XAS) spectrum of \UAl.
The top and right panels in Fig.~\ref{UAl3_RPES} (a) represent the photoemission spectrum measured at \hn{=725} and the \orb{U}{4d_{5/2}} XAS spectrum, respectively.
The XAS spectrum has a maximum at \hn{\sim 736.9}, and the photon energy of \hn{=725} is about $12~\mathrm{eV}$ below from the absorption energy.
The density plot in the center of Fig.~\ref{UAl3_RPES} (a) represents the difference between the spectrum measured at \hn{=725} and that measured at each photon energy.
The horizontal and vertical axes are the binding energy and the incident photon energy, respectively. 
All spectra are normalized to the maximum of the spectrum measured at \hn{=725} as a unity.

As the photon energy approaches to the \orb{U}{4d_{5/2}} absorption edge, the photoemission intensity just below \EF is enhanced.
The enhancement of the \Uf signal is much weaker than that observed in the \orb{Ce}{4f} signals at the \orb{Ce}{3d} absorption edge of $\mathrm{Ce}$-based compounds, where the enhancement factor is higher than 40 \cite{SF_review_JPCM}.
Nevertheless, the enhancement of \Uf signal coincides with the intensity of the XAS spectrum, and no $NVV$ Auger signal, which would appear as a diagonally-right down traces in the density plot, was observed.
This result indicates that the enhancement is indeed due to the Coster--Kronig type excitation in the \Udf resonant process, and not to the overlap of normal Auger signals.
Furthermore, a similar weak enhancement was observed at the \orb{U}{4d_{3/2}} absorption edge (\hn{\sim 778}, not shown), suggesting the enhancement originates from \Udf resonant process.

In Fig.~\ref{UAl3_RPES} (b), the on-resonance (\hn{=737}) and off-resonance (\hn{=732}) spectra, along with the corresponding difference spectrum, are shown.
The shape of the difference spectrum was found to be similar to the that of the valence band spectra of itinerant uranium compounds such as $\mathrm{UB}_2$ \cite{UB2_ARPES} and $\mathrm{UN}$ \cite{UN_ARPES}, and it is consistent with the itinerant \Uf nature of \UAl observed in our previous angle-resolved photoelectron spectroscopy (ARPES) study \cite{UAl3_ARPES}.

\subsection{Localized compound \UGa}
\begin{figure}[t]
	\includegraphics[scale=0.5]{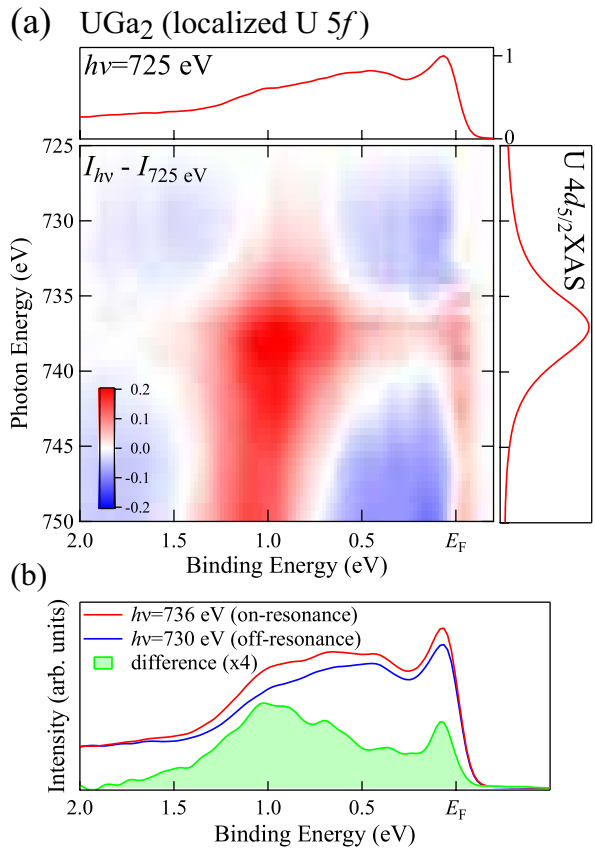}
	\caption{(Online color)
		RPES spectra of \UGa.
		(a) Density plot of RPES spectra together with the \orb{U}{4d_{5/2}} XAS spectrum.
		(b) On- and off-resonance spectra measured at $h \nu = 736$ and $730~\mathrm{eV}$, respectively, and the corresponding difference spectrum.
	}
\label{UGa2_RPES}
\end{figure}
Figure~\ref{UGa2_RPES} shows the same representation of the localized \Uf compound \UGa.
The spectrum measured at \hn{= 725}, which is shown in the top panel, consists of a sharp peak at the Fermi level and multiple peaks on the higher binding energy side.
In the photon energy dependence of the spectra shown in the central density plot, an enhancement is recognized although its appearance is very different from that of \UAl.
Two different energy locations of the resonance enhancement were observed: One at the Fermi level and the other centered at \EB{\sim 1}.
Since the enhancement in the latter is greater, the \Uf states are mainly localized in this compound.
Figure~\ref{UGa2_RPES} (b) shows the on-resonance (\hn{=736}) and off-resonance (\hn{=730}) spectra, and the corresponding difference spectrum.
The difference spectrum exhibits a sharp peak at the Fermi level and a broad peak centered at \EB{\sim 1} with a much stronger contribution, and its overall structure differs remarkably from that of \UAl.

\subsection{Heavy Fermion compound \UPdAl}
\begin{figure}[t]
	\includegraphics[scale=0.5]{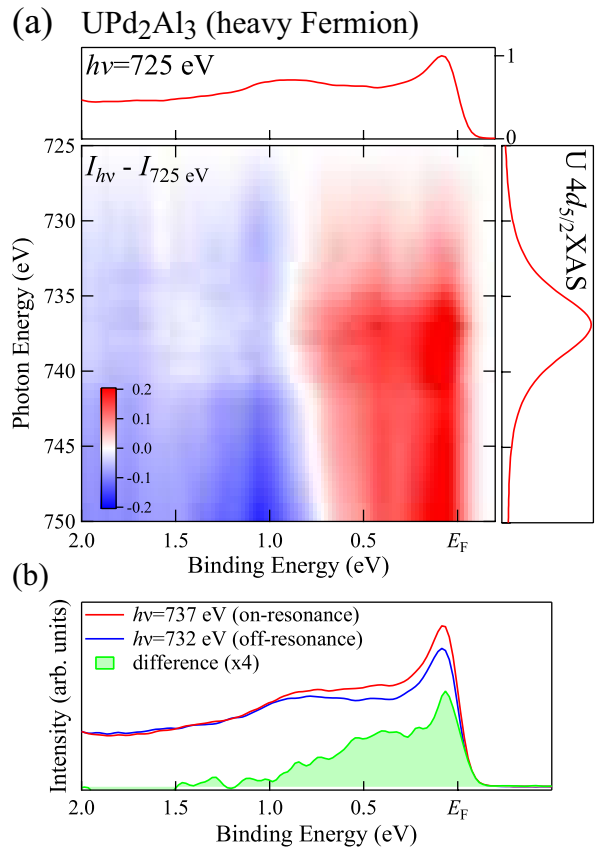}
	\caption{(Online color)
		RPES spectra of \UPdAl.
		(a) Density plot of RPES spectra together with the \orb{U}{4d_{5/2}} XAS spectrum.
		(b) On- and off-resonance spectra measured at $h \nu = 737$ and $732~\mathrm{eV}$, respectively, and the corresponding \Uf difference spectrum.
	}
\label{UPd2Al3_RPES}
\end{figure}
Figure~\ref{UPd2Al3_RPES} summarizes the \Udf RPES spectra of the heavy fermion compound \UPdAl.
The photon energy dependence of the spectra is different from the cases of \UAl and \UGa.
The intensity in the energy region of \EB{\lesssim 0.8} exhibits an enhancement at the \orb{U}{4d_{5/2}} absorption edge.
Furthermore, two vertical streaks are recognized in this image: One at the Fermi level and the other centered at \EB{\sim 0.4}.
Figure~\ref{UPd2Al3_RPES} (b) shows the on- and off-resonance spectra of \UPdAl measured at $h \nu = 737$ and $732~\mathrm{eV}$, respectively.
The difference spectrum is also indicated, and its profile is different from those of the itinerant compound \UAl and the localized compound \UGa.
The spectrum has a sharp peak at \EF, but in contrast with the spectrum of \UAl, there is a broad hump at \EB{\sim 0.4}.
Furthermore, its intensity is much weaker than that of the broad peak in the analogous spectrum of \UGa.

\subsection{Comparison with band-structure calculation}
\begin{figure}[t]
	\includegraphics[scale=0.5]{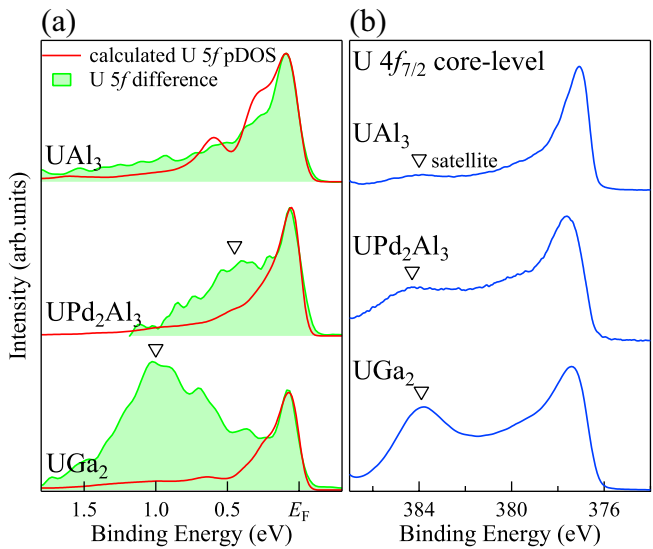}
	\caption{(Online color)
		(a) Comparison between the \Uf difference spectra of \UAl, \UPdAl, and \UGa and the \Uf pDOS from the band-structure calculation.
		Approximate positions of the incoherent component are indicated by inverted triangles in the spectra of \UPdAl and \UGa.
		(b) \orb{U}{4f_{7/2}} core-level spectra of \UAl, \UPdAl, and \UGa.
		Approximate positions of the satellite are indicated by inverted triangles.
	}
\label{U5fDOS}
\end{figure}
To further understand the implication of these \Uf difference spectra, we compared them with the calculated \Uf pDOS as shown in Fig.~\ref{U5fDOS} (a).
The red curves represent the \Uf pDOS obtained by the band-structure calculation based on the local density approximation (LDA) where all \Uf electrons are treated as itinerant.
The calculated \Uf pDOS were multiplied by the Fermi-Dirac function and broadened by the instrumental energy resolution to simulate the experimental \Uf difference spectra.
A systematic deviation of the calculated \Uf pDOS from the experimental \Uf difference spectra is recognized.
In the case of the itinerant compound \UAl, there is a good agreement between them.
Both of them have a sharp peak at the Fermi level and exhibit a long tail toward higher-binding energies.
On the other hand, in the case of the heavy fermion compound \UPdAl, although the peak at the Fermi level is well reproduced by the calculation, there is a broad peak around \EB{\sim 0.4} that cannot be explained by the calculation.
Furthermore, in the case of the localized compound \UGa, the intensity of the broad peak is remarkably enhanced, and its energy position is shifted toward higher-binding energies (\EB{\sim 1}). 
The structure is completely missing in the calculated \Uf pDOS.
Accordingly, the broad peak appears in the higher-binding energies in the valence band spectra of the heavy fermion compound and localized compound, and it cannot be explained within the framework of the LDA, suggesting that these broad peaks originate from the partially localized nature of \Uf states.

The partially localized nature of the \Uf states in these compounds were also observed in their core-level spectra which are a sensitive probe of the local electronic structures of uranium site \cite{Fujimori_SSC}.
Figure~\ref{U5fDOS} (b) shows the \orb{U}{4f_{7/2}} core-level spectra of these compounds.
Data were replotted from Refs. \cite{SF_review_JPSJ} and \cite{Ucore}.
In all spectra, the main peak is accompanied by a satellite at an approximately $7~\mathrm{eV}$ higher binding energy side of the main line.
This is designated as ``$7~\mathrm{eV}$ satellite'' \cite{Laub_Sat} which originates from the unscreened $\mathrm{U}~5f^{2}$ state in the photoemission final state \cite{SF_review_JPSJ,Ucore}.
Its intensity exhibits a similar behavior to that of the broad peaks in their valence band spectra: As the degree of the localization of \Uf state increases, the intensity of the satellite is enhanced. 
Thus, it should be reasonable to assume that the broad peaks in the valence band spectra also originate from a similar local-type excitation with the same unscreened $\mathrm{U}~5f^{2}$-dominant final state character.

The behavior of the double-peak structure of the \Uf pDOS coincides with that of the spectral profile of the Mott metal-insulator transition where the incoherent satellite peak is shifted toward higher binding energies and its intensity is enhanced as $U/W$ increases \cite{DMFT_Kotliar} ($U$ and $W$ are the on-site Coulomb energy and the one-electron band width, respectively).
Thus, the broad peak in the \Uf valence band spectra corresponds to the incoherent localized state with the $5f^2$ final state character, and the Hubbard-$U$ type mechanism takes an essential role in the $5f$ electronic structure.

Here, note that the previous ARPES studies on \UPdAl revealed that the bands at the Fermi level are renormalized due to the electron correlation effect in the energy scale of less than few-hundred meV \cite{UPd2Al3_ARPES1, UPd2Al3_ARPES2, UM2Al3_ARPES, SF_review_JPSJ}, but presumably the angle-integrated nature hindered their detection in these difference spectra.
Thus the correlation effect in \Uf states appears in two different energy scales:
Renormalization of bands in the vicinity of the Fermi level \cite{USb2_kink} and the appearance of the incoherent peak on higher binding energies.
This hierarchal nature of the electron structure in \Uf compounds was also theoretically predicted by DMFT+$U$ calculation and the intermediate Coulomb-$U$ coupling \cite{Das_review}.

\begin{figure}[t]
	\includegraphics[scale=0.5]{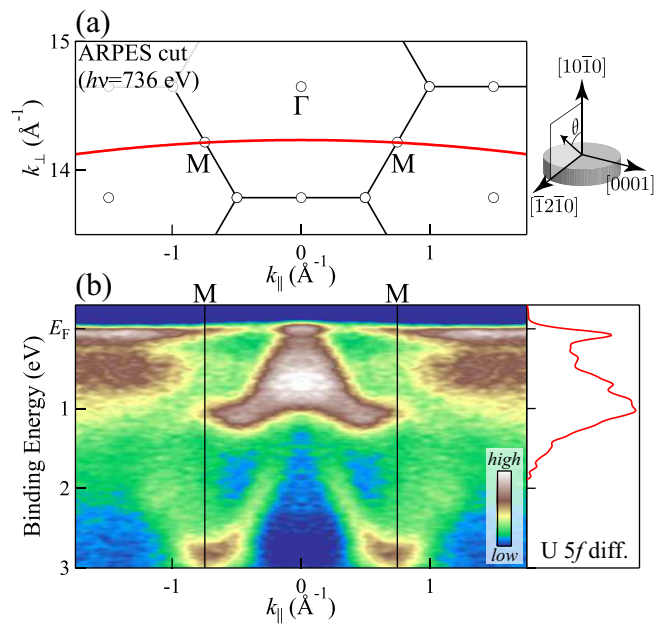}
	\caption{(Online color)
		ARPES spectra of \UGa measured at \hn{=736}, which is the on-resonant condition.
		(a) ARPES cut in the momentum space.
		(b) ARPES spectra together with the \Uf difference spectrum.
	}
\label{UGa2_ARPES}
\end{figure}
\subsection{Resonant ARPES study of \UGa}
To further unveil the nature of the incoherent peak, we have measured the ARPES spectra of \UGa at \hn{=736} which corresponds to the on-resonance condition.
In the experimental setup, the sample surface was parallel to the $[10\overline{1}0]$ axis, and the angular scan was along $[\overline{1}2\overline{1}0]$ direction.
The ARPES cut traces in momentum space along the $\mathrm{M-M}$ direction as illustrated in Fig.~\ref{UGa2_ARPES} (a).
Figure~\ref{UGa2_ARPES} (b) shows the ARPES spectra of \UGa measured along the ARPES cut together with the \Uf difference spectrum in the left panel.
In addition to the narrow band at the Fermi level, dispersive bands were observed in the energy region of \EB{=0.2 - 1.2}, where the incoherent \Uf peak has a dominant contribution in the \Uf difference spectrum.
In particular, a bell-shaped structure with an energy dispersion of about $1~\mathrm{eV}$ was observed around the $\mathrm{\Gamma}$ point, suggesting that the incoherent ``localized'' state also have a sizable hybridization with ligand states.

The dispersive nature of the incoherent peak in $5f$ compounds is in accord with the theoretical calculations \cite{Das_review,Zwicknagl_UPd2Al3}.
Similar dispersive nature of the incoherent peak was also reported in the transition metal $3d$ compound $\mathrm{SrVO_3}$ \cite{SrVO3}, and thus it could be a common feature of incoherent states.
On the other hand, the energy dispersions were hardly observed experimentally in the incoherent peak of the heavy fermion compound $\mathrm{CeIrIn}_5$ \cite{CeIrIn5_ARPES_Chen}, suggesting that there exist fundamental differences in the energy scale between $\mathrm{Ce}$ and $\mathrm{U}$ compounds although their transport properties are often very similar each other.
Such fundamentally-different natures between $5f$ and $4f$ states were also reported by the recent systematic analysis of ferromagnetic $5f$ compounds \cite{Tateiwa_UFM}.

\section{CONCLUSION}
In summary, we have revealed the \Uf electronic structures of typical uranium compounds using \Udf RPES.
Incoherent satellite peaks were observed in the \Uf spectra of the heavy fermion compound \UPdAl and the localized compound \UGa, whose behavior fit well with the mechanism of the Mott metal-insulator type transition.
Moreover, the unique physical properties such as unconventional superconductivity emerge in the intermediate Coulomb $U$ region as in the case of \UPdAl.
The incoherent component of \UGa exhibits an energy dispersion of about $1 \mathrm{eV}$, suggesting that it has a sizable hybridization.
These results indicate that the nature of the localized $5f$ electrons differs from that of $4f$ compounds, and the application of a simple localized model is not suitable for their description even for localized \Uf compounds.

\acknowledgments
The authors thank A. Tanaka, M. Taguchi, and K. Okada for stimulating discussion.
The experiment was performed under Proposal Nos 2014B3820, 2015A3820, 2015B3820, 2016A3810, and 2016B3811 at SPring-8 BL23SU.
The present work was financially supported by JSPS KAKENHI Grant Numbers 26400374 and 16H01084 (J-Physics).

\bibliographystyle{apsrev}
\bibliography{URPES}

\end{document}